\documentclass{article}

\usepackage{microtype}
\usepackage{graphicx}
\usepackage{subfigure}
\usepackage{booktabs} 

\usepackage{hyperref}

\usepackage{multirow}

\usepackage[accepted]{icml2025}

\usepackage{amsmath}
\usepackage{amssymb}
\usepackage{mathtools}
\usepackage{amsthm}

\usepackage[capitalize,noabbrev]{cleveref}

\theoremstyle{plain}

\theoremstyle{definition}

\theoremstyle{remark}

\usepackage[textsize=tiny]{todonotes}

\icmltitlerunning{Towards Distributed Inference of LLMs on a P2P Network}

\begin{document}

\twocolumn[
\icmltitle{Towards Distributed Inference of LLMs on a P2P Network}



\icmlsetsymbol{equal}{*}

\begin{icmlauthorlist}
\icmlauthor{Shabari S Nair}{ut,equal}
\icmlauthor{Krishanu Saini}{ut,equal}
\end{icmlauthorlist}

\icmlaffiliation{ut}{The University of Texas at Austin}

\icmlcorrespondingauthor{Krishanu Saini}{krishanu.saini@utexas.edu}
\icmlcorrespondingauthor{Shabari S Nair}{shabarisnair@utexas.edu}

\icmlkeywords{Machine Learning, LLM, Distributed Inference, Peer to Peer Networks}

\vskip 0.3in
]


\icmlaffiliation{ut}{Department of Computer Science, The University of Texas at Austin}




\printAffiliationsAndNotice{\icmlEqualContribution \quad
  \textit{Preliminary preprint.}}

\begin{abstract}
Prefix caching can reduce LLM inference latency by reusing KV caches across requests with shared prompts, but cluster-scale reuse is challenging because caches are partitioned across nodes. We propose a decentralized, prefix-cache-aware routing scheme for peer-to-peer LLM serving. Each node maintains a local radix tree of its own cached prefixes and asynchronously refreshed estimates of peer caches using periodic anti-entropy. Requests are routed to the node with the longest estimated prefix match, without centralized coordination or KV-cache transfer. Stale metadata only causes cache misses, not incorrect outputs, making weak consistency sufficient for correctness. Evaluation on simulated MMLU workloads show that decentralized routing improves latency under low communication delay and skewed prefix distributions, while high network latency and affinity-induced hotspots limit its benefits.
\end{abstract}

\section{Introduction}

Large language model (LLM) inference is increasingly constrained not only by model size, but also by redundant computation across requests. In autoregressive generation, each request first performs a \emph{prefill} phase over the prompt tokens, producing key--value (KV) cache entries at every Transformer layer. These KV states depend only on the input prefix and are subsequently reused during decoding. Consequently, when multiple requests share common prefixes---for example, shared system prompts, few-shot demonstrations, tool-use instructions, agentic scaffolds, or repeated conversation histories---the corresponding KV caches are identical over the shared prefix. Recomputing these states for every request wastes GPU cycles, increases time-to-first-token latency, and reduces serving throughput.

Prefix caching addresses this redundancy by retaining KV cache entries after a request completes and reusing them when a later request has a matching prefix. Prior serving systems have shown that this optimization can substantially reduce prefill cost. In vLLM, PagedAttention enables efficient KV-cache memory management and supports hash-based reuse of cached blocks \cite{kwon2023vllm}. In SGLang, prefix reuse is organized using a radix tree that maps token prefixes to cached KV states, enabling efficient lookup, sharing, and eviction of common prompt prefixes \cite{zheng2024sglang}. In a single-node setting, prefix caching is therefore largely a local data-structure and memory-management problem.

However, at the cluster scale, prefix caching becomes a distributed-systems problem. In a deployment with multiple LLM replicas, each worker maintains its own GPU-resident KV cache. These caches are naturally partitioned across nodes: a prefix cached on one worker is not automatically available on another. A request routed to a worker that does not hold its prefix must recompute the entire prefill, even if the same prefix exists elsewhere in the cluster. Thus, the effectiveness of prefix caching depends critically on the routing layer. A cache-aware router must direct each request to a worker likely to contain the relevant prefix, while also preserving load balance, avoiding hotspots, and tolerating stale or incomplete cache metadata.

Existing systems typically address this challenge using either centralized routing or shared cache storage. Centralized routers, as in cluster-level cache-aware serving designs, maintain a global or near-global view of worker cache state and dispatch requests accordingly \cite{zheng2024sglang}. While effective, this architecture introduces a logically centralized coordination point that can become a scalability bottleneck, a source of additional latency, and a potential availability concern under partial failures. Alternatively, shared KV-cache systems such as Mooncake and LMCache allow workers to fetch cached KV states from a shared cache pool or remote memory layer \cite{qin2025mooncake, liu2025lmcache}. These systems are powerful in datacenter environments, but transferring KV caches over the network can be expensive: even modest prefixes may occupy tens or hundreds of megabytes, and larger models can produce gigabyte-scale KV states for long contexts. As a result, remote KV transfer is not always suitable for bandwidth-constrained, heterogeneous, edge, or peer-to-peer deployments.

In this paper, we study an alternative design point: \emph{fully decentralized, prefix-cache-aware routing for distributed LLM inference}. We consider a peer-to-peer cluster of LLM serving nodes in which every node both serves requests and participates in routing. Each node maintains a local radix tree representing the prefixes currently cached in its own GPU memory. In addition, each node maintains approximate, read-only replicas of peer radix trees. These peer trie estimates are refreshed asynchronously using periodic anti-entropy: nodes periodically exchange compact cache-state summaries, allowing their views of the cluster to converge over time without requiring synchronous coordination.

This design turns cache-aware routing into an eventually consistent metadata-replication problem. Each incoming request is routed independently by the node that receives it, using only that node's local view of the cluster. If the local view indicates that a peer is likely to contain a longer matching prefix, the request may be forwarded to that peer; otherwise, it is served locally. Crucially, stale routing metadata affects only performance, not correctness. If a node routes a request based on stale trie information, the selected worker may simply miss the cache and recompute the prefill. The generated output remains correct because every worker still runs the full LLM. Thus, unlike replicated databases or consensus protocols, cache-aware routing does not require strong consistency for safety. It is a natural fit for \emph{coordination avoidance}: the system avoids distributed agreement in the steady state because inconsistency only causes lost optimization opportunities rather than semantic violations \cite{bailis2014coordination}.

Our design is inspired by classical eventually consistent systems. Periodic anti-entropy has long been used to reconcile divergent replicas, including in Bayou \cite{terry1995managing} and Dynamo-style storage systems \cite{decandia2007dynamo}. However, our setting differs from traditional replicated storage in two important ways. First, each trie has a single writer: the node that owns the corresponding KV cache. Other nodes only maintain read-only approximations of that trie. This avoids write-write conflicts and eliminates the need for consensus or conflict resolution. Second, the replicated state is advisory rather than authoritative. Trie replicas are used only to improve routing decisions; they do not determine the correctness of model execution. This makes weak consistency especially attractive: cache metadata can be stale, incomplete, or temporarily unavailable without compromising the functional behavior of inference.

The resulting architecture is compelling for several reasons. First, it removes the centralized routing tier, allowing the system to scale and degrade more gracefully under partial failure. Second, it avoids transferring large KV tensors across the network; instead, only lightweight cache metadata is exchanged. Third, it composes naturally with existing single-node prefix caching mechanisms, such as radix-tree-based cache lookup. Fourth, it is well-suited to peer-to-peer, edge, or multi-router deployments where requests may enter the system at many locations and where a single global scheduler is undesirable. Finally, it exposes a rich distributed-systems trade-off between locality, load balancing, metadata staleness, routing overhead, and cache eviction.

\paragraph{Contributions: }
This paper makes the following contributions:

\begin{itemize}
    \item We propose a fully decentralized architecture for prefix-cache-aware distributed LLM inference. In our design, every serving node runs identical routing logic and makes local routing decisions without relying on a centralized coordinator.

    \item We introduce an approximate radix-tree replication mechanism for cluster-wide cache awareness. Each node maintains an exact radix tree for its own KV cache and asynchronously refreshed, read-only estimates of peer cache states using periodic anti-entropy.

    \item We frame cache-aware LLM routing as a coordination-avoidance problem. We show that weakly consistent cache metadata is sufficient because stale routing decisions affect only performance through cache misses, not the correctness of LLM generation.

    \item We evaluate the proposed routing scheme on a simulated four-node LLM serving cluster using a prefill-heavy MMLU workload. Our experiments study latency reductions from prefix reuse, cache-hit behavior, and the emergence of specialization--eviction cycles under pure affinity-based routing.

    \item We analyze the fault-tolerance properties of the design under crash failures, stale trie replicas, lost metadata broadcasts, and failed forwarded requests. We argue that the system requires no explicit recovery protocol for correctness, because failures only effect liveness rather than violating inference quality.
\end{itemize}

Overall, this work explores a new point in the design space of LLM serving systems: decentralized, weakly consistent, locality-aware inference routing. Rather than treating distributed prefix caching as a problem requiring centralized scheduling or remote KV-cache transfer, we investigate whether lightweight metadata replication and blocking coordination-free peer-to-peer routing can recover much of the benefit of prefix reuse while preserving scalability, availability, and deployment flexibility.

\section{System Design}


Consider $N$ nodes $(P_1,P_2...P_N)$, each running an LLM able to serve requests. We assume all the LLMs pertain to the same underlying model, operating on the same GPU hardware. Each node $P_i$ has a specified amount of GPU memory, CPU memory, and local storage. The problem is given an incident request to run inference on an input on one the nodes, how can it best serve this request by utilizing other nodes in the network. We operate in a setting of crash-only failure (no Byzantine behavior), and disregard any privacy concerns that might arise in the practical application of such a P2P network.

\begin{figure}
    \centering
    \includegraphics[width=1.0\linewidth]{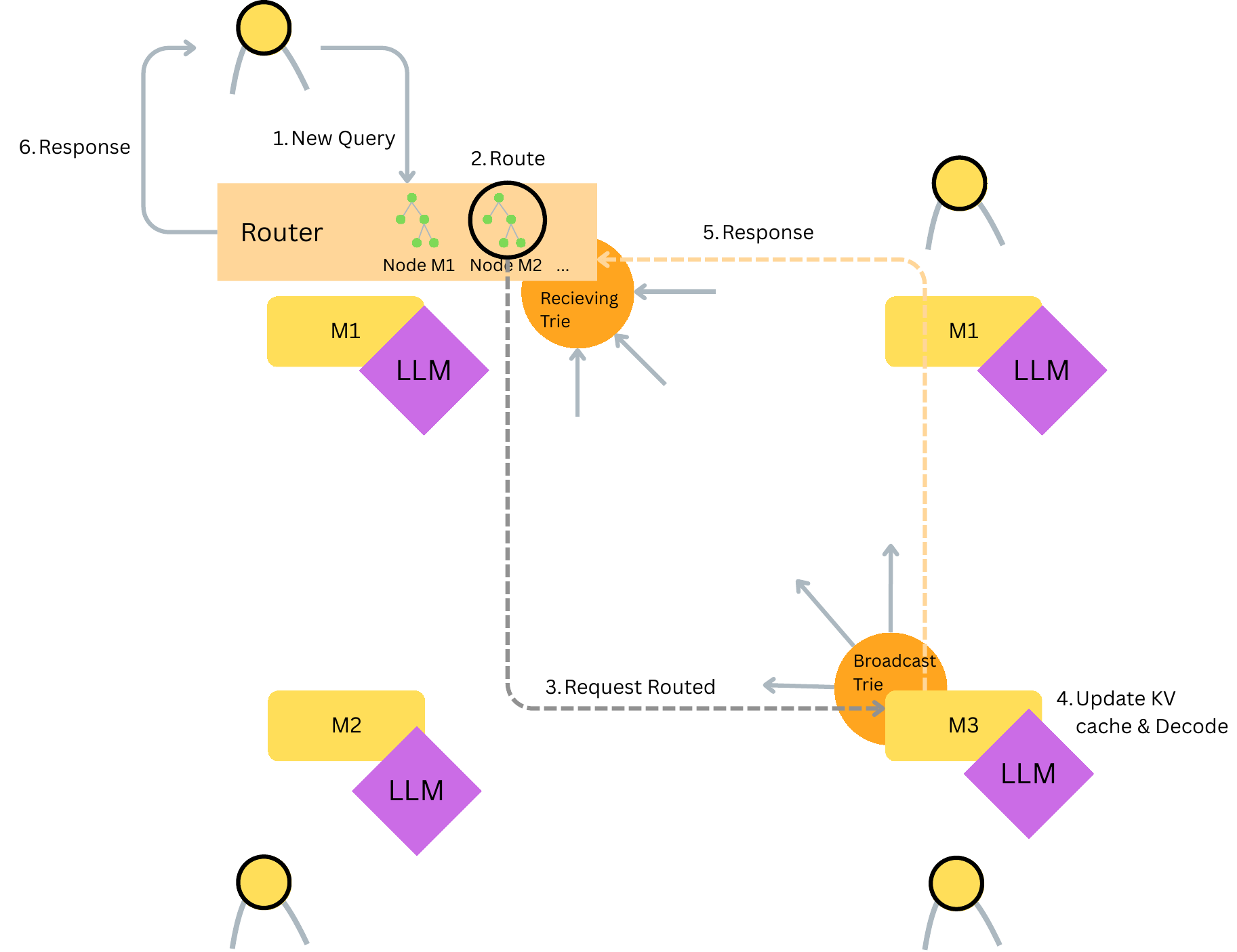}
    \caption{Flow for a new incoming user request arriving at node M1. Each node has a local router that keeps lazily updated information of the prefixes processed by different nodes. The query gets routed to the best node using a cache-aware criterion. The response is processed and returned to the correspondence node.}
    \label{fig:arch}
\end{figure}

We consider solving this problem through efficient KV cache sharing between nodes. Basically, we propose using a Radix Tree which is an optimized trie data structure, similar to SGLang, for each node to maintain a record of prefixes whose KV cache it has currently in its memory, as well as what it believes is stored by other nodes in the network. This would lead to each node storing one such trie for itself, as well as $N-1$ for the others. Then when a particular node receives an input request, it can decide which node to reroute the request to based on the largest prefix match between the input request and Tries of the remaining nodes. The end-to-end flow is shown in Figure \ref{fig:arch}. Our algorithm considers each node having access to the following primitives

\begin{itemize}

\item \texttt{runInfer(req)}: Executes inference for request \texttt{req} using the local LLM hosted on node $P_i$.

\item \texttt{getBestNode(req)}: Uses node $P_i$'s estimates of the prefix tries of other nodes to identify the node $k$ whose trie has the largest prefix match with \texttt{req}.

\item \texttt{updateTrie(req)}: Inserts request \texttt{req} into node $P_i$'s local prefix trie.

\item \texttt{updateTrieEstimate(req, k)}: Updates node $P_i$'s estimate of node $P_k$'s trie by inserting the request \texttt{req}.

\item \texttt{updateKVCache(req)}: Adds any missing KV cache entries corresponding to request \texttt{req} into the local GPU memory.

\item \texttt{evictLRUTrie()}: Removes one least-recently-used leaf token from the local trie and returns the evicted token.

\item \texttt{evictKVCache(tok)}: Evicts the KV cache associated with token \texttt{tok}, thereby freeing GPU memory.

\item \texttt{sendTrie()}: Broadcasts node $P_i$'s local trie structure to all other nodes in the network.

\item \texttt{sendRequest(req, k)}: Sends request \texttt{req} to node $P_k$ for processing and returns a success or failure response.

\end{itemize}

We provide the pseudo-code for the procedure for handling requests in Fig \ref{main_algo}. Note that for this method to work, each node would have to share its trie structure with all the other nodes (\textbf{anti-entropy}). One way to achieve this is to havdo it to have nodes periodically (based on local clocks) share their trie structures with others in the background. This is non-blocking since it happens 'asynchronously' with other processes, like making routing decisions or request processing. This achieves \textbf{eventual consistency}, wherein in the absence of new requests, all nodes would eventually have an up-to-date estimate of other nodes' KV caches. Another way could be an 'on-demand' approach, wherein a node requests this information from all the other nodes as and when it requires. However, while this would result in up-to-date estimates of KV caches of other nodes for every node prior to processing any request or making any routing decision, this then becomes a 'blocking' call since it would prevent a new request from being processed or routed till all fresh trie structures of other nodes have been obtained, thus increasing latency. We make the design choice of prioritising latency reduction at the cost of consistency, and hence choose the former non-blocking method of updating tries. Practically, we implement this using a \textbf{gossip-protocol}.

An orthogonal approach to our method of sharing prefix tries is to share entire KV caches between nodes. This would result in a more direct use of KV caches of other nodes to update your own, following which processing can be done within the node that received the request. However, this is generally impractical as, unlike characters, KV caches are collections of vectors. As such, they occupy a much higher memory and are not conducive to being transported across arbitraty distances in arbitrary time intervals on a possibly low bandwidth medium. For context, memory of KV cache for the Qwen2.5-0.5B model is around 14KB per token, which amounts to around 42MB for a modest request of 3000 tokens. This cost only grows with model size and prefix length. Forwarding the request rather than the cache shifts this cost off the critical path and lets the system tolerate network conditions that a cache-fetch design could not.

\begin{algorithm}[H]
\caption{Handling an Incoming Request $r$ at Node $P_i$: $\textsc{SendRequest}(r,i)$}
\label{main_algo}
\begin{algorithmic}[1]

\STATE \textbf{Initialize:}
\STATE $r \gets \text{incoming request}$
\STATE $Trie_i \gets \text{radix tree of node } i$
\STATE $\widehat{Trie}_j \gets \text{estimate of node } j\text{'s radix tree at node } i$

\STATE $k \gets \arg\max\limits_{j \in \{1,\dots,N\}} \textsc{PrefixMatch}(r,\widehat{Trie}_j)$

\IF{$k = i$}
    \WHILE{required memory $>$ available GPU memory}
        \STATE $tok \gets \textsc{EvictLRUTrie}()$
        \STATE $\textsc{EvictKVCache}(tok)$
    \ENDWHILE

    \STATE $(result, kv\_cache) \gets \textsc{RunInfer}(r)$
    \STATE $\textsc{UpdateKVCache}(kv\_cache)$
    \STATE $\textsc{UpdateTrie}(result)$
\ELSE
    \STATE $ack \gets \textsc{SendRequest}(r,k)$

    \IF{$ack.status = \text{success}$}
        \STATE $\textsc{UpdateTrieEstimate}(r,k)$
    \ENDIF
\ENDIF

\STATE \textbf{return} $result$

\end{algorithmic}
\end{algorithm}




    












We also note that the above procedure works under a \textbf{crash recovery model} for the LLM nodes, wherein nodes may fail and come back online later on. When a node fails, its KV cache is not persisted and is instead completely lost. This does not violate the safety constraints as requests can be processed without any KV cache (just slower).

\section{Safety Conditions and Failure Handling}
\label{sec:failure-handling}

Here we discuss the safety and liveness conditions for our system, along with failure and possible performance degradation cases that can arise.

\subsection{Asynchrony}

The design's safety properties hold under fully asynchronous execution. Routing decisions are made entirely from local state and produce a result in bounded time regardless of network conditions; the single-writer invariant on each trie $T_i$ is enforced by construction; and cache state converges eventually whenever broadcasts are eventually delivered. None of these properties require any assumption on message delay or processing time. The system as a whole assumes partial synchrony in the sense of Dwork et al.~\cite{dwork1988consensus}, but only for \emph{liveness}: anti-entropy convergence requires that broadcasts are eventually delivered, and detecting a crashed peer during request forwarding requires that timeouts eventually fire. Also note that safety holds even if our timeout-based error detector is inaccurate. In this case, the system falls back to local execution using a timeout. 

This minimal synchrony requirement is itself a consequence of the system's coordination-avoidance design~\cite{bailis2014coordination}. Because routing decisions never block on inter-node agreement, no consensus or quorum protocol is invoked, and the synchrony assumptions those protocols would require are absent. The single property the design relies on for correctness --- that stale routing information produces cache misses rather than incorrect outputs --- is preserved regardless of network behavior: a request routed to a peer whose trie estimate is arbitrarily out of date will still yield a correct inference result, with only the prefill cost re-incurred. 

\subsection{Hot specialist push-back}

Pure prefix-affinity routing creates a structural risk: if a node $P_k$ accumulates a popular prefix in its radix cache, every request matching that prefix will be forwarded to $P_k$, regardless of $P_k$'s current load. The result is queueing at the ``specialist'' node while peers sit idle --- a load-imbalance pathology intrinsic to affinity-only routing. We address this with a \textbf{push-back} mechanism similar to network congestion control \cite{ioannidis2002implementing}: when $P_k$'s queue depth exceeds a threshold, it advertises this in the next anti-entropy broadcast alongside its trie. Originating nodes incorporate the back-pressure signal into their routing decisions, pausing routing requests to $P_k$ until a specified time period, by when its load is expected to have subsided. Push-back does not require any new coordination --- it piggybacks on the existing broadcast channel --- and does not change the safety properties of routing: a request that would otherwise go to an overloaded specialist is simply executed locally or sent to a slightly inferior at a partial-cache cost, which is the same fallback behavior the system already exhibits under crash failures.

\subsection{In-flight forwarded request failure}

If $P_i$ forwards a request to $P_k$ and $P_k$ crashes before responding, $P_i$ must detect the failure and recover. We use a timeout-based detector co-located with the anti-entropy channel: the broadcast that ships $P_i$'s trie also serves as a heartbeat from $P_i$, so the absence of broadcasts from $P_k$ within a multiple of the broadcast interval $\Delta_{poll}$ marks $P_k$ as suspected-failed. On a \texttt{sendRequest} timeout, the originating node falls back to the next-best routing target --- which could be itself --- and runs the request on that node. This recovery requires partial synchrony only for liveness (the timeout must eventually fire on a real crash); safety holds under arbitrary asynchrony because the originating node always retains the option of local execution.

\subsection{Trie-broadcast loss}

A dropped or delayed broadcast simply leaves peers' estimates stale for an additional interval. Because every broadcast carries a complete snapshot of the sender's local trie --- not an incremental delta --- no log replay or message reordering is required to recover: convergence resumes at the next successfully delivered round. The system tolerates arbitrary loss patterns short of a permanent partition, and even under a permanent partition, the partitioned subsets continue to make correct local routing decisions on stale information, with the only consequence being increased cache misses. This is the structural payoff of using anti-entropy with full snapshots rather than a logged update protocol. Note that this mechanism is similar to state-based CRDTs, and as such it also has the capability to tolerate duplicated and missed trie broadcast messages.

\subsection{Stale trie estimates}

Even in the absence of broadcast loss, a peer's estimate $\widehat{T}_k$ at any moment lags $P_k$'s actual trie $T_k$ by up to one broadcast interval. If $P_i$ routes to $P_k$ based on an entry that $P_k$ has since evicted, $P_k$ must recompute the prefix; the routing decision pays the cost of a remote forward without the benefit of a cache hit. We do not detect or correct staleness explicitly. Instead, the design tolerates it because the consequence --- a cache miss --- is a performance event, not a correctness violation, and the next anti-entropy round restores convergence. The staleness budget is bounded above by $\Delta_{poll}$ in expectation, and we quantify its effect on routing quality empirically in Section~\ref{sec:results}.

\section{Experimental Setup}
\label{sec:experimental_setup}

All our experiments make use of LLM nodes running as separate processes, with simulated communication delays. The network send/receive latency has a fixed component with a small random jitter, assuming these costs are uniform across all nodes. All experiments consists of a cluster of four nodes running on independent partitions of a single H100 GPU, each alloted 10\% of the total GPU VRAM. Each node has an LLM inference backend component implemented via SGLang, and a routing layer for handling trie sharing and processing logic. Here we do not interfere with how SGLang handles cache eviction and updation logic.

We use the MMLU dataset, wherein we use 5 question/answer pairs from each topic to construct the few-shot examples for all the remaining questions from that topic. This serves to create a very prefill-heavy dataset, especially since we limit the decoding to one output token, since the answer to each question is a multiple choice option. All the requests are then randomly arranged in a random schedule with the requests distributed between the four nodes in a round robin fashion. Each request is sent one by one, only after the previous request has been completed. This results in batching-free evaluation. Note that we are not concerned with the quality of output generated (which remains the same), and are only interested in the improvement in efficiency. 

To test sensitivity, we also simulate an alternate skewed request in which 80\% of requests come from 20\% of topics. This is a Zipfian-style skew approximating real-world conversational workloads where some prompts dominate.

\section{Results and Discussion}
\label{sec:results}

\paragraph{Latency Analysis: } 
The end-to-end latency of a request broadly decomposes into four components: the cost of SGLang inference (prefill and decode), the I/O cost of radix-tree management, request processing overhead, and the blocking communication cost incurred when a request is forwarded to a remote node. To characterize the economic break-even of forwarding, we ablate Communication Latency $\Delta_{comm}$ across \{2\,ms, 200\,ms, 1000\,ms\}, corresponding to LAN, cross-region, and unstable network conditions. The break-even point follows directly from a simple model: forwarding is profitable when the prefill time saved by reusing a remote cache entry exceeds the round-trip cost of the forward. Also note that the anti-entropy step is non-blocking so it's not included in the critical-path analysis.
\begin{table}[t]
\centering
\small
\caption{Mean per-request latency across configurations, with paired no-routing baselines.}
\label{tab:latency}
\begin{tabular}{lrrr}
\toprule
\textbf{$\Delta_{comm}$} & \textbf{Routed (s)} & \textbf{No-route (s)} & \textbf{Gap (s)} \\
\midrule
2\,ms     & 0.310 & 0.323 & $-0.013$ \\
200\,ms   & 0.785 & 0.323 & $+0.462$ \\
1000\,ms  & 2.271 & 0.323 & $+1.948$ \\
\midrule
\multicolumn{4}{l}{\textit{Workload skew ($\Delta_{comm} = 2$\,ms, $\Delta_{\text{poll}} = 500$\,ms)}} \\
\midrule
skewed    & 0.206 & 0.266 & $-0.060$ \\
\bottomrule
\end{tabular}
\end{table}

\begin{figure}
    \centering
    \includegraphics[width=0.9\linewidth]{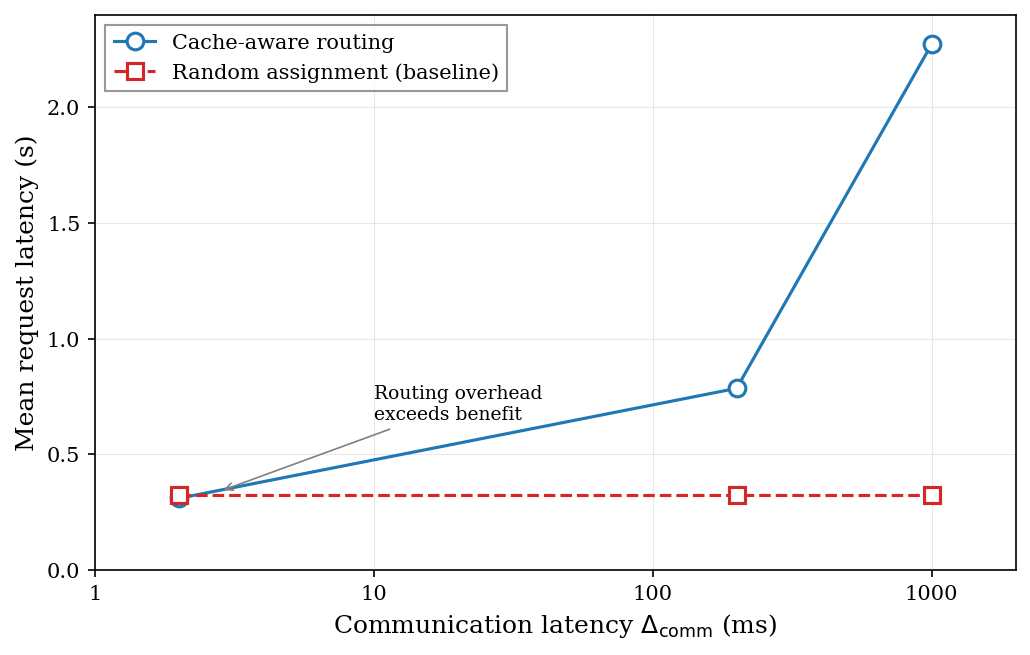}
    \caption{Mean per-request inference latency vs $\Delta_{comm}$ comparison under decentralized routing versus a no-routing baseline on general dataset.}
    \label{fig:placeholder}
\end{figure}

From Table~\ref{tab:latency}, the break-even point lies between 2\,ms and 200\,ms of communication latency: at 2\,ms, routing wins on cache-hit-heavy workloads, while at 200\,ms the round-trip cost exceeds the prefill savings. Note that skewed settings has a larger gap of $0.060$, showing more benefit of decentralized routing in specialized settings. Our experiments use a 0.5B-parameter model with a mean prompt length of $\sim$540 tokens; larger models and longer prompts both increase the per-token prefill cost, pushing the break-even point higher and expanding the regime in which decentralized routing is profitable.

\paragraph{Cache-Hit Analysis: }
The cache-hit rate offers insight into both node specialization and routing effectiveness, ultimately reflected in end-to-end latency. From table \ref{tab:workload-comparison}, in non-routed baselines, the hit ratio comes in around $83$--$85\%$. This is expected because the prefix is common across each topic, accounting for roughly five-sixths of the whole prefill. Against this baseline, routing produces only a $0.7\%$ improvement under uniform workloads, a marginal gain. This is reasonable given our experimental scale, in which a small dataset and abundant GPU memory allow each node to retain essentially all relevant KV-cache state regardless of routing. Under skewed workloads, however, routing yields a $6.7\%$ hit-rate improvement, attributable to the formation of specialist nodes that accumulate deep caches for the dominant prefixes. This contrast confirms that decentralized routing's cache benefit is workload-dependent.

\begin{table}[t]
\centering
\small
\caption{Comparison of decentralized routing versus a no-routing baseline under uniform and skewed workloads ($\Delta_{\text{comm}} = 2$\,ms, $\Delta_{\text{poll}} = 500$\,ms). Expected cache hit rate is the routing node's prediction at decision time; actual hit rate is measured at the processing node.}
\label{tab:workload-comparison}
\begin{tabular}{lrrr}
\toprule
\textbf{Configuration} & \textbf{Exp. Hit (\%)} & \textbf{Act. Hit (\%)} & \textbf{Lat (s)} \\
\midrule
Uniform, no-route & 47.3 & 83.1 & 0.323  \\
Uniform, routed   & 66.4 & 83.7 & 0.310  \\
\midrule
skewed, no-route & 67.4 & 85.6 & 0.266   \\
skewed, routed   & \textbf{88.6} & \textbf{91.4} & \textbf{0.206} \\
\bottomrule
\end{tabular}
\end{table}

\begin{figure*}
    \centering
    \includegraphics[width=0.9\textwidth]{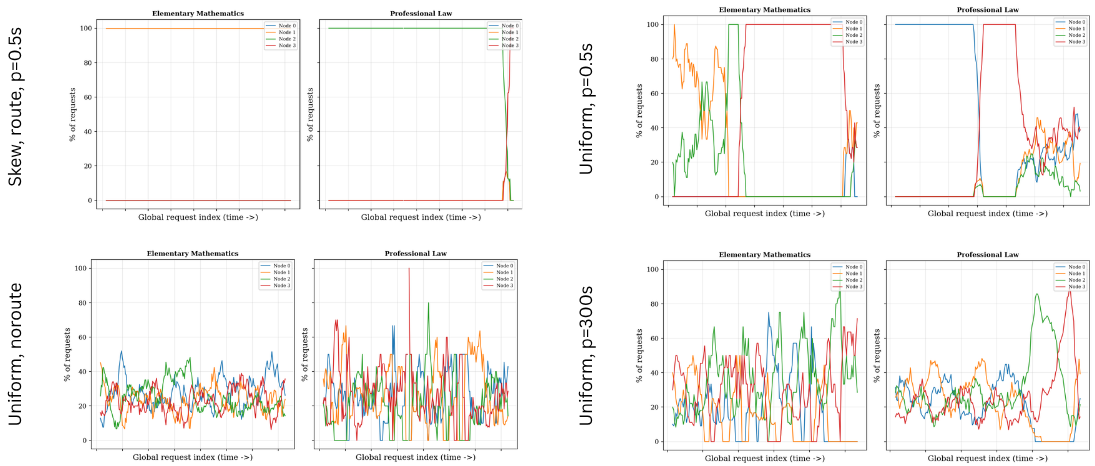}
\caption{Per-node request distribution over time for two MMLU subjects under four configurations. Here, the y-axis indicates the percentage of requests processed by a node, averaged across a window of 300 past requests, moving with a stride of 30. Routing produces sharp specialization under skew (top-left) and the specialization--eviction cycle under uniform load (top-right); no-routing yields the $\sim$25\% uniform baseline (bottom-left); slow polling ($\Delta_{\text{poll}}=300$\,s) weakens specialization as peer estimates go stale (bottom-right).}    
\label{fig:routing_all}
\end{figure*}

\begin{figure}[!ht]
    \centering
    \includegraphics[width=1.0\linewidth]{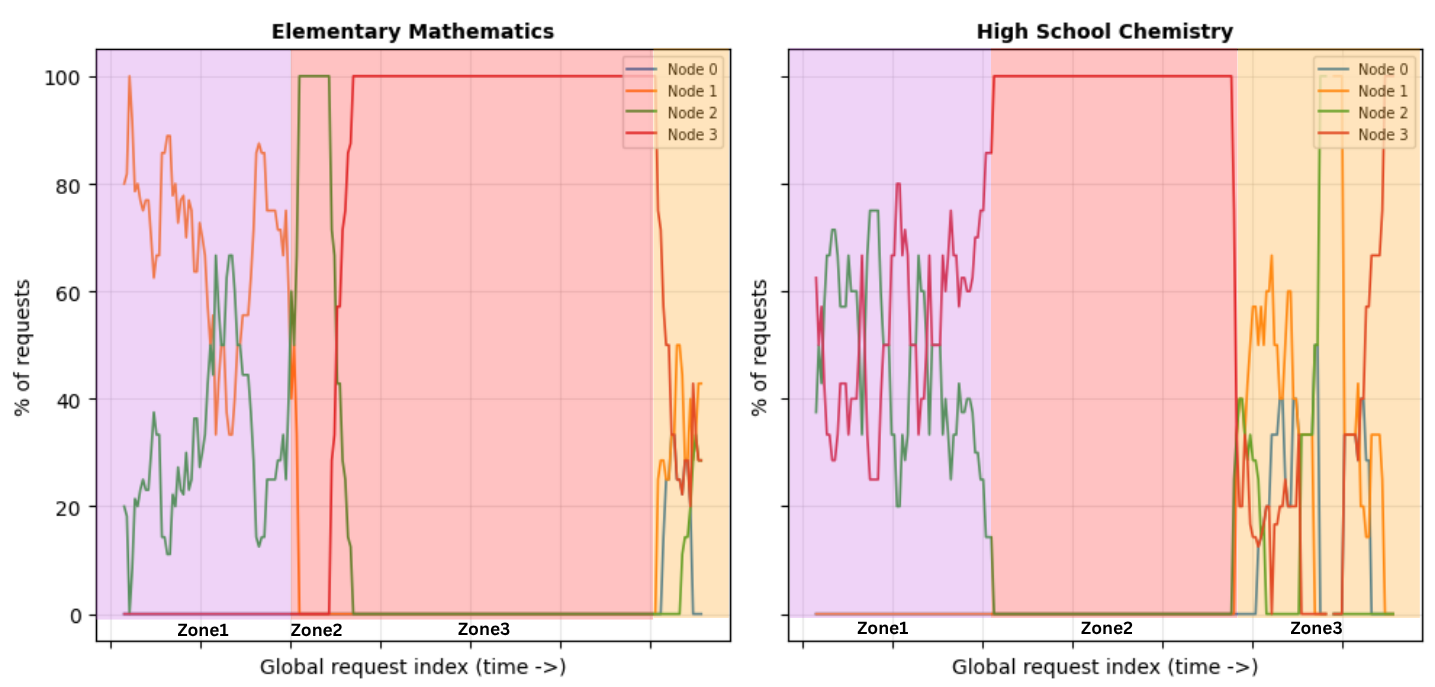}
    \caption{Per-node share of requests over time for two MMLU subjects, showing the three-phase specialization--eviction cycle: a competitive phase (Zone~1) with no clear specialist, a specialization phase (Zone~2) where one node captures essentially $100\%$ of subject traffic, and an eviction phase (Zone~3) where LRU pressure collapses the hotspot, and traffic redisperses. Here, the y-axis indicates the percentage of requests processed by a node, averaged across a window of 300 past requests, moving with a stride of 30.}
    \label{specialization-cycle}
\end{figure}

\paragraph{Specialization–eviction cycle: }
We analyze the topic-wise request routing patterns over time in different configurations. Specifically, we want to see the emergence of specialization on topics due to the accumulation of cache, the imbalance caused by the eviction of caches, and equilibrium states. We demonstrate an experiment with configuration: $\Delta_{comm} = 200ms$, $\Delta_{poll} = 500ms$ in Figure \ref{specialization-cycle} illustrating the specialization-eviction cycle for two topics. Each graph shows the share of requests routed to each of the four nodes. Three regimes are visible: an initial competitive phase (Zone~1) in which no node holds the topic's prefix and traffic is dispersed across nodes, a specialization phase (Zone~2) in which one node captures essentially 100\% of the topic's requests once its trie accumulates a dominant prefix match, and an eviction phase (Zone~3) in which LRU eviction displaces the cached prefix and traffic re-disperses. The same structure recurs across topics.

\paragraph{Quantifying Hot-spots: }
Prefix Cache-based routing concentrates traffic on whichever node first accumulates a popular prefix, producing the specialization--eviction behavious described above. To quantify this concentration, we measure the share of requests captured by the single most-loaded node in each run. Under a perfectly uniform data distribution across our four nodes, this metric should yield $25\%$, however, the no-routing baseline reaches $28.7\%$ under uniform workloads and $34.7\%$ under skewed workloads, slightly above uniform due to natural variance in random request assignment. Our decentralized routing framework pushes this concentration to $65$--$76\%$ under uniform workloads (Table~\ref{tab:forwarding-hotspots}) and to $88\%$ under skewed workloads --- a $2.3$--$2.6\times$ amplification of imbalance attributable purely to the routing logic. The intensity of this concentration is modulated by the anti-entropy frequency: as polling slows from $20$\,ms to $300$\,s, peak load falls from $74.8\%$ to $42.2\%$, because peers operating on stale views distribute requests without knowing the current cache state.


\begin{table}[t]
\centering
\small
\caption{Hotspot concentration across configurations, measured as the share of requests routed to the single most-loaded node.}
\label{tab:forwarding-hotspots}
\begin{tabular}{lr}
\toprule
\textbf{Configuration} & \textbf{Hotspot (\%)} \\
\midrule
\midrule
\multicolumn{2}{l}{\textit{No-routing baseline}} \\
\midrule
uniform workload, no-route       & 28.7 \\
skewed workload, no-route        & 34.7 \\
\midrule
\multicolumn{2}{l}{\textit{Communication latency sweep ($\Delta_{\text{poll}} = 500$\,ms)}} \\
\midrule
comm 1\,s, routed                & 64.9 \\
comm 200\,ms, routed             & 67.4 \\
comm 2\,ms, routed               & 70.0 \\
\midrule
\multicolumn{2}{l}{\textit{Polling period sweep ($\Delta_{comm}$ $2\pm0.5$\,ms)}} \\
\midrule
poll 20\,ms, routed              & 74.8 \\
poll 100\,ms, routed             & 75.7 \\
poll 2\,s, routed                & 66.9 \\
poll 300\,s, routed              & 42.2 \\
\midrule
\multicolumn{2}{l}{\textit{Workload skew ($\Delta_{comm}$ 2\,ms, $\Delta_{\text{poll}} = 500$\,ms)}} \\
\midrule
skewed, routed                   & 88.1 \\
\bottomrule
\end{tabular}
\end{table}

\section{Conclusion}
We have presented a decentralized prefix-cache-aware routing scheme for distributed LLM serving and characterized the regimes in which it pays off. Using MMLU as a controlled sandbox of short prompts with bounded topic diversity, we have shown that the design is most effective on prefill-heavy workloads with skewed prefix distributions, where routing-induced specialization concentrates popular prefixes deeply enough to overcome forwarding and queueing overhead. Larger models, longer prefixes, and lower-latency networks should each push the break-even point further in the design's favor, suggesting that decentralized routing can become competitive with centralized routers and shared cache pools in deployments where these conditions are met. More broadly, the P2P architecture eliminates the single point of failure inherent to centralized designs, retains safety under asynchronous execution, and degrades gracefully under crash failures without explicit recovery protocols. We note that privacy concerns inherent to peer-to-peer cache sharing are application-dependent and beyond the scope of this work.

\bibliography{example_paper}
\bibliographystyle{icml2025}

\end{document}